# Infrared photothermal heterodyne imaging in thermally thick medium for thermo-optic property characterization

Jordan Letessier[1,2], Abel Netter[1,2], Jérémie Maire[1,2], Stéphane Chevalier[1,2,3,*]

[1] Univ. Bordeaux, CNRS, Bordeaux INP, I2M, UMR 5295, F-33400, Talence, France
[2] Arts et Metiers Institute of Technology, CNRS, Bordeaux INP, I2M, UMR 5295, F-33400 Talence, France
[3] LIMMS/CNRS-IIS(UMI2820), The University of Tokyo, 4-6-1 Komaba, Meguro-ku, Tokyo, 153-8505, Japan

**Abstract**

Measuring temperature fields in semi-transparent media requires the knowledge of material thermo-optic properties. While the current techniques are well established for isothermal thin films, they remain unexplored for thick media where temperature gradients need to be considered. This study proposes to generalize the current methods of infrared heterodyne photothermal imaging to thick media by combining transmittance measurements and thermo-optic modeling to measure simultaneously the thermoabsorbance and thermoreflectance coefficients in semi-transparent media. In this work, we demonstrate the validity of our methodology followed by the measurements of the thermo-optic properties of several materials: soft polymers (polydimethylsiloxane) and borofloat glasses.

**Keywords:** Thermo-optic coefficients, heterodyne technique, transmission measurements, thermal modeling, semi-transparent media

**\*Corresponding author:**
Prof. S. Chevalier
Mechanical Engineering Institute (I2M) of Bordeaux (CNRS UMR 5295)
Arts et Métiers Institute or Technology
Esplanade des Arts et Métiers

33405 Talence Cédex, France
Email: stephane.chevalier@u-bordeaux.fr

**Main text**

Photothermal imaging techniques have been extensively employed since the 1980s, starting with the development of photothermal deflection measurements [1–3]. Among them, photothermal heterodyne imaging has emerged as an ultrasensitive method for probing optical absorption, mapping temperature fields, and estimating concentrations of particles and molecules in solution [4,5]. Many recent studies have demonstrated the versatility of infrared photothermal heterodyne imaging (IR-PHI) for both chemical and thermal characterization at the microscale. In biology, the method has been successfully applied to, for instance, chemical mapping of molecular distributions [5] and for live-cell imaging with submicron resolution [4]. A comprehensive overview of photothermal microscopy for single-particle detection and its applications in materials and biology is provided in the review by Adhikari et al. [6]. Beyond biological imaging, IR-PHI in reflection has been successfully employed for submicron imaging of cation heterogeneities [7], and for thermal property measurements such as thermal conductivity and heat capacity of thin films [8] and anisotropic nanolayered materials [9]. The technique has also been employed for nanoparticle detection, highlighting its sensitivity even at the nanoscale [10,11]. Additionally, photothermal deflection-based approaches such as the mirage effect have been used for characterizing thermal properties of thin films, including two-dimensional materials like graphene [12,13].

As evidenced in the previous paragraph, PHI techniques have been widely validated [14,11,15] using a large range of application. Most of these methods rely on a pump−probe mechanism, in which a pump beam serves as the heating source and a probe beam detects the local refractive index change induced by the heat release from sample absorption. In practice, temperature measurements via PHI requires careful calibration to obtain a linear relationship between the variation of the probe light intensity (either in reflection or transmission) and the material local temperature [16]. Calibration of the technique relies heavily on the knowledge of the thermo-optic properties, whose characterization has been widely reported in recent years due to their growing importance in electronic devices for communication and sensing technologies [17–20]. Most studies have focused on characterizing these properties at a single wavelength and examining their evolution with temperature [17,21]. The thermo-optic coefficient describes the variation of the refractive index, $\tilde{n} = n + ik$, with temperature, $T$, such as $d\tilde{n}/dT$. For most materials (solid and liquid), these variations are generally very small, typically below $10^{-3}$ K$^{-1}$, with a typical order of magnitude around, e.g., $10^{-5}$ K$^{-1}$ in near infrared (IR) for 4H-SiC or GaN [17]. They can be measured experimentally using several techniques, such as interferometry with a Fabry-Perot interferometer, the z-scan technique, and thermal lensing [17,19–22]. All these methods generally simplify the physics by considering only the variation of the real part, $dn/dT$, in the case of reflection, or the imaginary part, $dk/dT$, in the case of absorption [23–25], or considering homogeneous temperature field in the sample. However, in the case of thermally thick layers, the presence of a thermal gradient does not allow to make these model simplifications, and the measurement of thermo-optic coefficients needs to be reconsidered.

Thus, this study addresses this gap in the literature and proposes an experimental setup based on PHI, but in transmission (which we call TPHI in this work), to measure thermo-optic coefficients. We move further toward a general thermo-optical model to quantitatively identify the variations in both reflectance ($dn/dT$) and absorbance ($dk/dT$) with temperature [26]. This method is applicable to a wide range of wavelengths and can be used for any semi-transparent media, enhancing its range of application since both thermoabsorbance and thermoreflectance vary with wavelength. To demonstrate its versatility in spectroscopy at microscale using camera as mid IR detector (with a spatial resolution of 23 µm/px), our approach is based on an IR laser pump (with a wavelength of $\lambda_{las} = 3.85$ µm) and use a non coherent broadband IR probe source ($\lambda_{probe} \in [2 - 11]$ µm to measure the thermo-optic coefficients in this range. Our work is the first milestone to realize 2D temperature field measurements in thick semi-transparent media where both thermoreflectance and thermoabsorbance happen. In the following sections, the method is presented, followed by the details of the thermo-optic and thermal modelling. Finally, the results of the characterization performed in different materials, i.e., borofloat glasses and polydimethylsiloxane (PDMS) are shown to demonstrate the validity of our methodology. A comparison with a thin layer of water is also presented.

The TPHI setup used in this study is similar to the transmission configuration shown in [11,27], but uses a camera as detector (X6580sc IR camera, InSb detector). Figure 1 describes the experimental setup used in our study. The blue path corresponds to the probe path using an IR broadband source emitting between 2 µm and 11 µm (IR source Thorlabs HPIR104). The red path corresponds to the pump path with the IR laser (laser Thorlabs QF3850T1 $\lambda_{las} = 3.85$ µm). A bandpass filter (Thorlabs FB4260-105 for PDMS, FB3250-500 for Borofloat glass, and FB5330-250 for Water) is used before the camera to eliminate the residual pump laser beam that is not absorbed and to control the wavelength at which the thermo-optic coefficient is measured.

To isolate the signal of interest from ambient radiation and the sample's own emission, we use the heterodyne technique [4,5]. This approach involves modulating both the thermal excitation (pump) and the probe source. By post-processing the signal captured by the camera, the relevant information is extracted at the sum and difference of the pump and probe frequencies. The pump and probe modulations are chosen to avoid harmonic relationships (non-multiple), in order to prevent any unwanted contributions in the heterodyne signals. The probe is modulated at the frequency, $f_s$, between 1 and 1.6 Hz, depending on the pump frequency, $f_{las}$, which is modulated between 0.1 and 0.625 Hz. This allows us to measure the variation in transmittance with different amplitudes of the temperature fields. Thermotransmittance images are acquired by the camera and spans at least 20 periods of the pump signal. Additionally, the total number of recorded images is set to be an integer multiple of both the pump and probe periods to ensure accurate Fast Fourier Transform. This corresponds to a camera acquisition rate between 20 and 48 Hz.

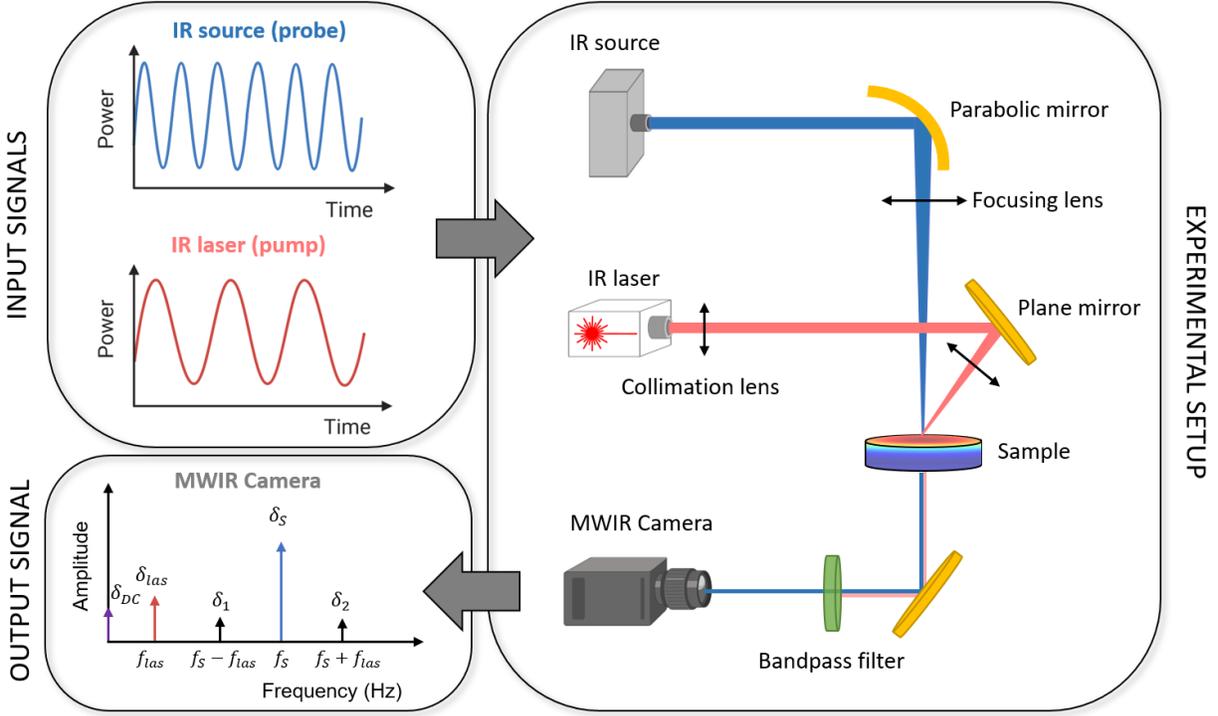

*Figure 1. Schematic of the experimental setup. The probe path is represented in blue and the heating path is in red. Input temporal signal are represented on the top left. The output signal is represented by its frequency spectrum.*

For each pixel of the images, the Fourier transform of the temporal signal is computed and a spectrum similar to the schematic shown in the output signal box in Figure 1 is obtained. The complex values, $\delta_1$ and $\delta_2$, at the heterodyne frequencies and at the probe frequency, $\delta_S$, are extracted to compute the thermotransmittance as follows

$$\frac{|\Delta\hat{\Gamma}|}{\Gamma_0}(x,y) = \frac{|\delta_1(x,y)| + |\delta_2(x,y)|}{|\delta_s(x,y)|}, \qquad 1.$$

where $|\delta_1|$, $|\delta_2|$, $|\delta_s|$ are the amplitude (modulus) of peak harmonic at heterodyne frequencies ($f_s - f_{las}$ and $f_s + f_{las}$) and probe frequencies ($f_s$). At a given pixel position, the modulus of thermotransmittance oscillations is given by $|\Delta\hat{\Gamma}|$. $\Gamma_0$ stands for the steady state thermotransmittance at the average temperature $T_0$. Therefore $|\Delta\Gamma|/\Gamma_0$ is the dimensionless modulus of thermotransmittance, and lies in the range from $10^{-2}$ to $10^{-4}$ K$^{-1}$ for the materials used in this study.

To be compared with the 1D thermal model described later, the thermotransmittance field is averaged over a region surrounding the thermal spot. The size of this region is determined by selecting the largest possible radius from the center of the spot up to the amplitude of the peaks at heterodyne frequencies ($\delta_1$ and $\delta_2$ in Figure 1, output signal box) start to be drown into the noise (when SNR<1.5).

The transmittance of a semitransparent material is defined as $\Gamma_0(x, y)$. In the case where this medium experiences a temperature variation over time, i.e., $\Delta T(x, y, t)$, then the medium transmittance becomes $\Gamma(x, y, t)$. In a semi-transparent and thermally thick material, as schemed in Figure 2, this quantity depends on several factor. In the authors' previous works [26], it was demonstrated that the thermotransmittance depends on both the average temperature of the medium and its interface temperature. For a specific position $(x, y)$ and a specific time, $t$, the local thermotransmittance can be written as:

$$\frac{\Delta\Gamma}{\Gamma_0}(x, y, t) \approx -\frac{\rho_0 \kappa_R}{1 - \rho_0}\big(\Delta T(x, y, z = 0, t) + \Delta T(x, y, z = e, t)\big)$$
$$- \alpha_0 \kappa_A \int_0^e \Delta T(x, y, z, t) dz,$$

2.

where $\rho_0$ and $\alpha_0$ are the reflectance and absorbance of the medium, respectively; $\kappa_R$ and $\kappa_A$ are the thermoreflectance and thermoabsorbance coefficients, respectively; $\Delta T$ represents the temperature field at a specific location $(x, y)$; $z = 0$ and $z = e$ represent the optical interfaces; and $e$ represents the material thickness (see Figure 2 for the axis orientation and the definition of dimension). In Equation (2), the first term on the right side is related to the thermoreflectance, whereas the second term is related to the thermoabsorbance. If the medium is isothermal, as is usually assumed, then $\frac{1}{e}\int_0^e \Delta T(x, y, z)dz \approx \Delta T(x, y, z = 0) \approx \Delta T(x, y, z = e)$, and the proportionality between the thermotransmittance and temperature increase, as in $|\Delta\Gamma|/\Gamma_0(x, y, t) \approx \kappa \Delta T(x, y, t)$, is ensured. However, as demonstrated in equation (2), the thermotransmittance signal is not proportional to the temperature increase or to the laser power absorbed. Therefore, the challenge is to first calibrate the thermotransmittance in order to determine the material's temperature.

In order to simplify the thermo-optical model parameters, equation (2) can be rewritten to:

$$\frac{\Delta\Gamma}{\Gamma_0}(x, y, t) \approx \kappa_R \left(\Delta T_{S_1}(x, y, t) + \Delta T_{S_2}(x, y, t)\right) + e \, \kappa_A \, \Delta T_V(x, y, t), \qquad 3.$$

where $\kappa_R$ [K$^{-1}$] and $\kappa_A$ [K$^{-1}$.mm$^{-1}$] are the thermoreflectance and thermoabsorbance coefficients for a given wavelength $\lambda_{probe}$, and are proportional to the derivative of the refractive index $\tilde{n}$ with respect to the temperature $T$ ($\kappa_R \propto \frac{\partial n}{\partial T}$ and $\kappa_T \propto \frac{\partial k}{\partial T}$). $\Delta T_{S_1}$ stands for the variation of temperature at the front face ($z = 0$), $\Delta T_{S_2}$ stands for the variation at the rear face ($z = e$), and $\Delta T_V$ is the volume-averaged temperature variation across the thickness.

During the TPHI experiment the heat source, and therefore the heat field, is modulated by the pump laser. Hence, it is more convenient to use the Fourier transform of the temperature, defined as:

$$\theta(x, y, \omega_{las}) = \int_{-\infty}^{+\infty} \Delta T(x, y, t) e^{-2i\pi \omega_{las} t} dt, \qquad 4.$$

where $\omega_{las} = 2\pi f_{las}$. Finally, equation (3) can be rewritten in the frequency domain at the laser pumping frequency as follows

$$\frac{\Delta \hat{\Gamma}}{\Gamma_0}(x, y, \omega_{las}) \approx \kappa_R \left( \theta_{S_1}(x, y, \omega_{las}) + \theta_{S_2}(x, y, \omega_{las}) \right) + e\, \kappa_A\, \theta_V(x, y, \omega_{las}), \qquad 5.$$

where $\Delta \hat{\Gamma}$ is the Fourier transform of the thermotransmittance, and $\theta_S$ and $\theta_V$ are the Fourier transforms of the temperatures associated with the surfaces and the volume. Based on this expression (equation 5), calibrating the thermo-optic coefficients requires modeling of the temperature rise.

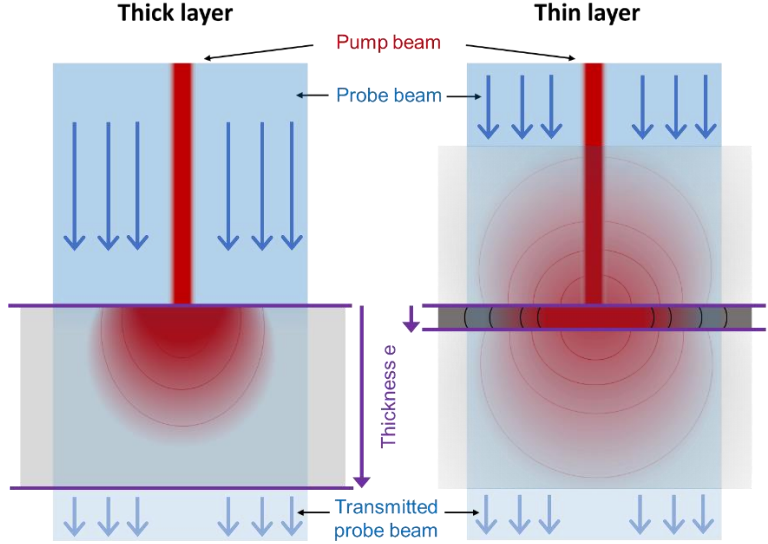

*Figure 2. Schematic of the experimental configuration for the thick layer (right) corresponding to the PDMS and Borofloat samples and for the thin isothermal layer (left) corresponding to the water sandwiched between two CaF2 layers,. The pump beam periodically heating the sample is shown in red, whereas the probe beam, shown in blue, is partially absorbed in the sample of interest of thickness e.*

A one-dimensional (1D) model is considered first in this study to eliminate the knowledge requirement of the exact shape of the heat excitation (which can be approximated by a gaussian profile with our laser heating). However, the total power deposited on the sample, $P_{las}$, is accurately measured using a power meter (Thorlabs S440C). By integrating the signal over a region that encompasses the entire energy deposited, the average heat distribution in a thick layer excited by a modulated laser is governed by the following equations:

$$\frac{d^2\theta}{dz^2}(z,\omega_{las}) - \gamma^2\theta(z,\omega_{las}) = -\frac{\mu\Delta\hat{P}_{las}(1-\rho_0)}{k}e^{-\mu z}, \qquad 6.$$

where $\gamma = \sqrt{i\omega/a}$ is the inverse of the complex characteristic diffusion length (unit m$^{-1}$), $a$ is the thermal diffusivity, $k$ is the thermal conductivity, $\Delta\hat{P}_{las}$ is the laser modulation amplitude, $\mu$ is the absorptivity coefficient, such as $\alpha_0 = e^{-\mu e}$, $i = \sqrt{-1}$ is the imagery unit, and $z$ the thickness dimension. The second term in the equation correspond to the volumetric heat source term. This source decays exponentially in depth due to Beer-Lambert absorption, i.e., $e^{-\mu z}$. To solve this equation, two boundary conditions are used, under the assumption of semi-infinite heat transfer in thick medium (true if the laser frequency is high enough). This latter assumption will be discussed in the results section. The boundary conditions are given as follows:

$$\frac{d\theta(z, \omega_{las})}{dz}\bigg|_{z=0} = 0, \qquad \qquad 7.$$

$$\theta(z \to \infty, \omega_{las}) = 0. \qquad \qquad 8.$$

In these conditions, one can find that the solution to this equation is given by:

$$\theta(z, \omega) = \frac{\mu \Delta \hat{P}_{las}(1 - \rho_0)}{k(\gamma^2 - \mu^2)} \left( e^{-\mu z} - \frac{\mu}{\gamma} e^{-\gamma z} \right). \qquad \qquad 9.$$

From this equation the complex temperature on the front face ($\theta_S$) is obtained as:

$$\theta_S(\omega) = \theta(0, \omega) = \frac{\mu}{\gamma} \frac{\Delta \hat{P}_{las}(1 - \rho_0)}{k(\mu + \gamma)}, \qquad \qquad 10.$$

and the average complex temperature in the volume ($\theta_V$) is obtained by integrating over z as:

$$\theta_V(\omega) = \frac{1}{e} \int_0^e \theta(z, \omega) dz = \frac{\Delta \hat{P}_{las}(1 - \rho_0)}{ek(\gamma^2 - \mu^2)} \left( 1 - e^{-\mu e} - \frac{\mu^2}{\gamma^2}(1 - e^{-\mu e}) \right), \qquad \qquad 11.$$

For semi-infinite samples, equation (5), (10) and (11), predict the variation of transmittance $\Delta \Gamma / \Gamma_0$ in case of semi-transparent media heated by laser modulated excitation. The required thermal parameters, conductivity and specific heat, were measured in-house by the transient plane source technique (Hot Disk® TPS2500S) and differential scanning calorimetry (Setaram DSC131), respectively, prior to carrying out the thermotransmittance measurements. The value obtained for PDMS and borofloat glasses are given in Table 1.

*Table 1. Thermo-optic parameters measured in-house for PDMS, Borofloat and CaF2 wafers.*

| Parameters | PDMS | Borofloat | CaF$_2$ |
|---|---|---|---|
| Volumetric mass density ($\rho$) (g.cm$^{-3}$) | 1.008 | 2.205 | 3.180 |
| Specific heat ($c_p$) (J.kg$^{-1}$.K$^{-1}$) | 1519 | 830 | 1167 |
| Thermal conductivity ($k$) (W.m$^{-1}$.K$^{-1}$) | 0.177 | 1.078 | 9.620 |
| Thermal diffusivity ($a$) (mm$^2$.s$^{-1}$) | 0.116 | 0.589 | 2.593 |
| Optical extinction coefficient ($\mu$) (mm$^{-1}$) | 0.2302 | 1.3700 | 0.0025 |
| Optical reflection coefficient. ($\rho$) (%) | 1.90 | 2.45 | 2.70 |

In order to extend the TPHI method to thick samples, we first use a thermally thin material to validate our approach. The configuration chosen comprises a thin layer of water sandwiched between two CaF$_2$ layers. Water was chosen for its strong absorption at the pump laser wavelength. PDMS and Borofloat glass were selected as the thick samples, whose thermo-optic coefficients are estimated and discussed below. Both configurations, thermally thin and thick samples, are schematically shown in Figure 2 with the heat distribution represented in red.

In the case of a thin layer heated by a modulated heat source, the absorbed heat diffuses into the neighbouring layers: the CaF2 wafer in this case (see schematic in Figure 2). Such configuration is well known, and it has already been shown [28,29] that the temperature of the thin layer can be well estimated by $\theta_{TL} = \Delta \hat{P}_{las}(1 - e^{-\mu e})(1 - \rho_0)/k\sqrt{a/i\omega}$ where $k$ and $a$ are the CaF$_2$ thermal conductivity and diffusivity, respectively, $\Delta \hat{P}_{las}$ is the amplitude of the modulated laser and $\mu$ is the water absorptivity at 3.85 µm. This expression is similar to [30], where the temperature increase of the thin layer is proportional to $f^{-1/2}$. Such behavior is characteristic of thermal transport in a thin layer deposited on substrate. Thermotransmittance measurement using TPHI were performed in three water layer thicknesses ($e$ = 0.05, 0.1 and 0.2 mm) in a range of laser frequencies ($f_{las}$ = 0.1 to 1 Hz)

The results presented in Figure 3(a) correspond to a thin water film ($e$ = 0.1 mm, $f_{las}$ = 0.4 Hz, $P_{las}$ = 0.12 W). They show the thermotransmittance maps, where each pixel is calculated according to equation 1. The center of the image, where the laser heats the water, is visible, and the heat diffuses over 100-200 µm in all directions around the spot. The maximum of the thermotransmittance amplitude is approximatively 0.27, with the expected order of magnitude for an elevation of around 10 degrees, given the laser power and an thermo-optic coefficient value ($\kappa$) of approximately 0.01 K$^{-1}$ [31].

The thermotransmittance signal was then averaged over an area defined by the region where the $SNR > 1.5$ around the laser spot. These average values were then plotted against the estimated average temperature of the water layer (calculated over the same area) for various frequencies and thicknesses. These results are presented in Figure 3(b). As mentioned earlier, in the case of a thermally thin layer deposited on a substrate, the thermo-optical model reduces to $\frac{\Delta \Gamma}{\Gamma_0} = \kappa \, \theta_{TL}$ with $\kappa = 2\kappa_R + \kappa_A e$. This relation is presented in Figure 3(b) with solid lines, confirming the linearity of our measurements. The differences of the slope are related to the thickness of the water layers, and they can be used to measure $\kappa_R$ and $\kappa_A$ in Figure 3(d). A linear regression in Figure 3(d) enables to identify $\kappa_R$ from the intercept and $\kappa_A$ from the slope, with a R value of 0.99. A last interesting point is the frequency behavior of the thin layer thermotransmittance observed in Figure 3(c). In these data, the average thermotransmittance is presented versus the laser frequency in log-log scale. As expected, a $-1/2$ slope is observed, validating our measurements, setup and methodology (see $\theta_{TL}$ equation above).

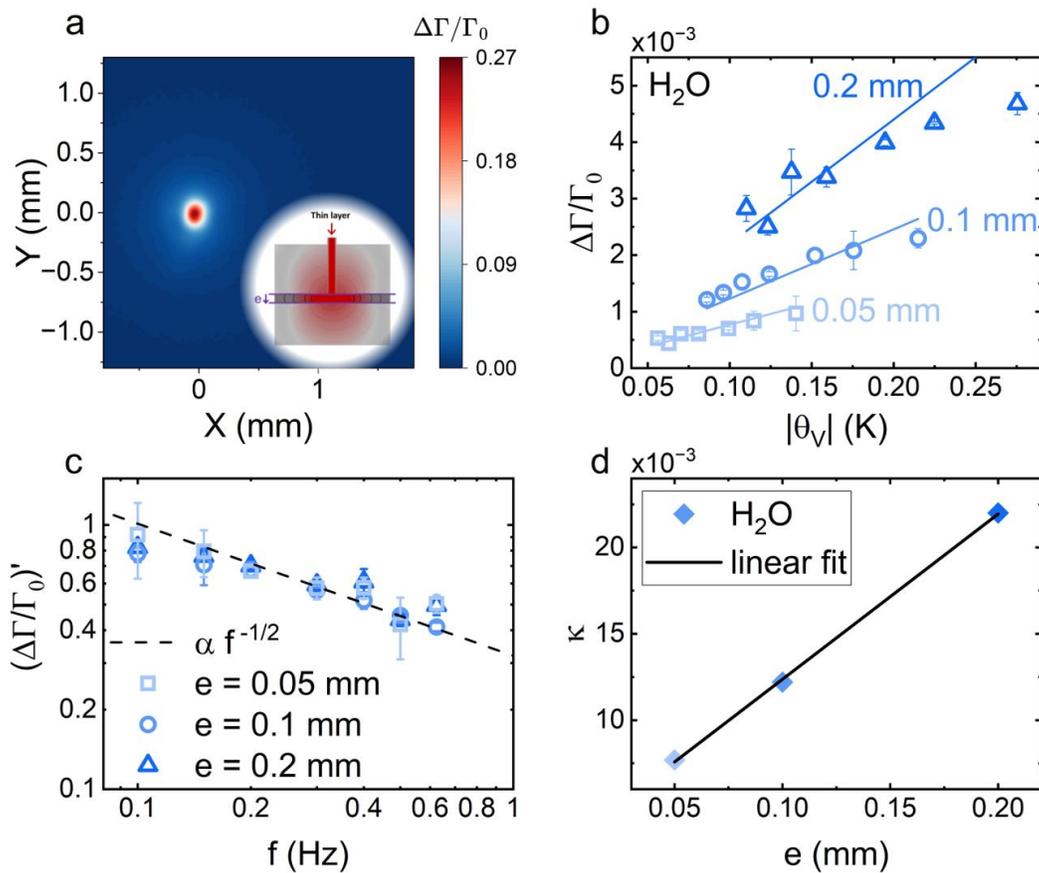

*Figure 3. (a) 2D map of the variation of transmittance ($\lambda_{probe} = 5.33$ μm) through 0.1mm of water heated by a modulated laser source (amplitude of 0.12 W at 0.4Hz). (b) Relative variation of transmittance through three thicknesses of water. 0.05 mm (squares), 0.1 mm (circles) and 0.2 mm (triangles). (c) Normalized relative variation of transmittance for isothermal water layers as a function of heating frequency. Normalization was performed with respect to the maximum of the linear regression of the data obtained at 0.2 Hz. The data follows a $f^{-1/2}$ trend. (d) Extensive coefficient of thermotransmittance for different thicknesses of water.*

The thermotransmittance obtained in thermally thick layers, i.e., in PDMS and Borofloat are presented in Figure 4(a) and (b). Both images present similar patterns with greater thermotransmittance at the center where the laser beam heats the sample. Thermal diffusion then spreads the heat from the central hotspot. It is important to note that the averaged thermotransmittance signal in PDMS is more than three times greater than in the Borofloat wafers despite the lower pump power. Following the same processing as in the previous section, the thermotransmittance signal was averaged over an area defined by the region where the $SNR > 1.5$ around the laser spot, which corresponds to a radius of 1.84 mm for the Borofloat and PDMS. The normalized modulus of this averaged thermotransmittance was then plotted versus the expected modulus of the material temperature over a frequency range in Figure 4(c).

For thick materials, transmittance variations follow the equation $|\kappa_R \theta_S(z=0,\omega) + \kappa_A e\, \theta_V(z,\omega)|$ (see equation 5). Both $\theta_S(z=0,\omega)$ and $\theta_V(z,\omega)$ scale as $f^{-1/2}$ at low frequencies and as $f^{-1}$ at high frequencies (see equation 10 and 11), defining the asymptotic behaviors represented by the dotted and solid lines. The data align with these expectations, validating our thermo-optical model. Such change in the scaling law is also a very good indicator of the presence of thermally thick materials, and it is advised to check before performing temperature or material chemistry measurement using TPHI. However, caution is needed at low frequencies, as the semi-infinite medium assumption may not remain satisfied given that the diffusive length becomes too large compared to the material thickness, which could also account for the observed change in behavior. This is not the case in our study as the material thickness remains sufficiently large to avoid such effects.

In comparison, in thin water films, the black dotted curve follows a $f^{-1/2}$ trend, characteristic of a thin thermal model, even at relatively high frequencies (higher than 0.25 Hz). The thin layer assumption may not hold true at very high frequencies, where the thermal boundary layer becomes comparable to the film thickness, which could be the case for measurements performed with a thickness of 0.2 mm. In Borofloat glass, the transition from low to high frequency behavior occurs around 0.2 Hz. For PDMS, the transition is more gradual, and the low-frequency trend is not yet fully visible, depending on the absorptivity µ. Since $\mu = 0.23$ mm$^{-1}$ for PDMS and 1.37 mm$^{-1}$ for Borofloat, the term $\mu\sqrt{\frac{i\omega}{a}}$ in the expression of $\theta_S$ becomes dominant over $\frac{i\omega}{a}$ at lower frequencies (see equation 12) for Borofloat.

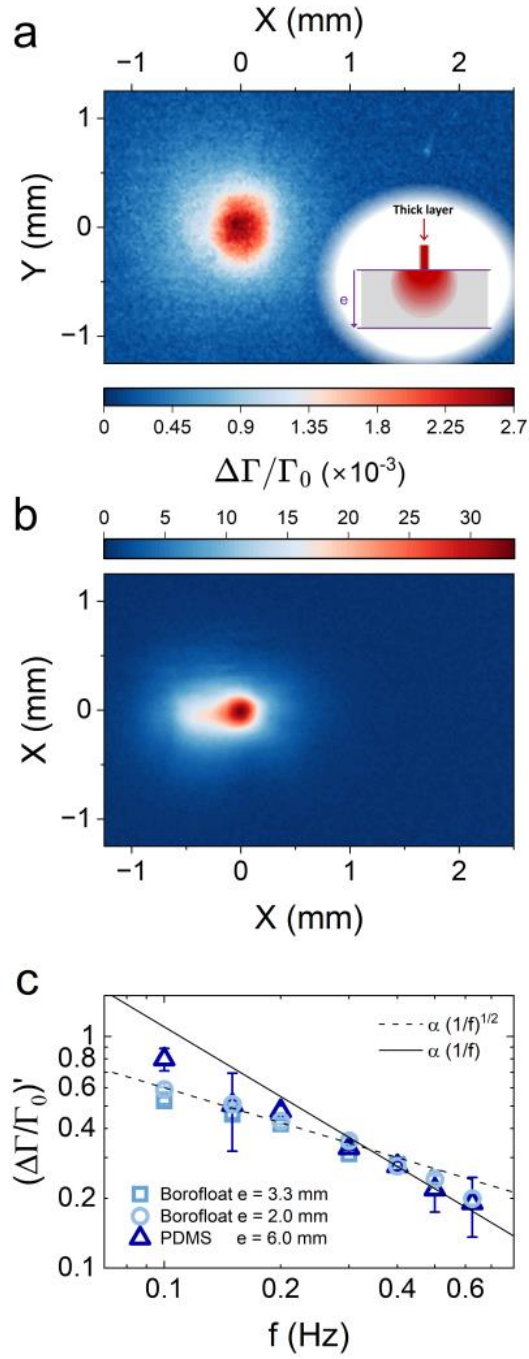

*Figure 4. (a) 2D map of the variation of transmittance ($\lambda_{probe} = 3{,}25$ μm) through 2.0mm of glass Borofloat heated by a modulated laser source (amplitude of 0.12 W at 0.4Hz). (b) Normalized variation of transmittance signal for thick layers of Borofloat and PDMS. Normalization was performed with respect to the maximum of the linear regression of the data obtained at 0.2 Hz. The data follows the trends in $f^{-1/2}$ at low frequency and $f^{-1}$ at high frequency.*

As mentioned earlier, the thermo-optic coefficients of thin layers were computed from the slope of the thermotransmittance measurements with respect to the averaged volumetric temperature $\theta_V$. Here, the same method was used with thick layers, albeit the relationship between the measurements and $\theta_V$ becomes non-linear. In this case, knowing the material thermal and optical properties, the laser power and its frequency, the calculation of the modulus of equation 5 leads to a nonlinear relationship between $\kappa_A$, $\kappa_R$ and $|\Delta\hat{\Gamma}|/\Gamma_0$, because surface temperature at the front face is not equal to the mean temperature ($|\theta_S(z=0)| \neq |\theta_V|$). Figure 5 clearly shows that the variation of the transmittance signal is no longer proportional to the mean temperature of the layer ($|\theta_V|$). Therefore, a non-linear least square algorithm was used to fit the best curves considering equation 5 and the thermal model in equation 11. For each curve, the value of $\kappa_A$ and $\kappa_R$ were estimated. The obtained values are given in the Table 2.

The values obtained are the first reported for PDMS material in the mid-IR range. The value obtained for Borofloat glass matches that previously obtained by the authors in a previous work using another method, thus further validating this approach [26]. As it can be seen in Figure 4, the thermotransmittance effect is significantly higher in PDMS than in Borofloat, leading to significantly higher PDMS thermo-optic coefficients. This high value obtained for PDMS suggests that TPHI measurements are highly sensitive in PDMS, much more than in the other materials investigated so far, which opens interesting applications in PDMS-based microfluidics or for ultra-low temperature increase detection. In this work, the temperature variation detected in PDMS at the edge of the averaging circle, i.e., where the SNR is equal to 1.5, is on the order of 30 mK. Finally, the thermo-optic coefficients obtained in thermally thick materials demonstrate that the effects of reflection and absorption must be considered simultaneously to fit the appropriate model. This provides additional insight for the use of TPHI in thermally thick media.

Pump-probe measurements are often performed in a regime in which the temperature increase generated by the pump does not significantly affect the thermal properties of the material being investigated. Here, we are confronted to this same limitation. For high pump power, significant variations of $\mu$ with temperature could be observed. In these conditions, the temperature increase is large enough to impact the thermal properties of the material, but the thermal model also becomes non-linear since the absorption coefficient cannot be considered constant anymore ($\mu = f(T)$). Therefore, measurements in materials with high thermo-optic coefficients such as PDMS, require low pump power. In the experiment presented in this work, the power used was $P_{las} = 22$ mW to ensure that $\mu$ could be considered constant.

The second limitation relates to wavelengths. Since the transmitted signal is measured, no transducer is used to absorb the pump laser energy. Hence care must be taken to choose the pump wavelength so that the material of interest absorbs the incoming energy. A wavelength-accordable laser is particularly well suited to this. On the contrary, the main requirement for the probe wavelength is that it is partially transmitted by the sample. Considering the importance of both wavelengths, measuring the full IR spectrum of the material prior to the TPHI measurement is advised to ensure optimal SNR.

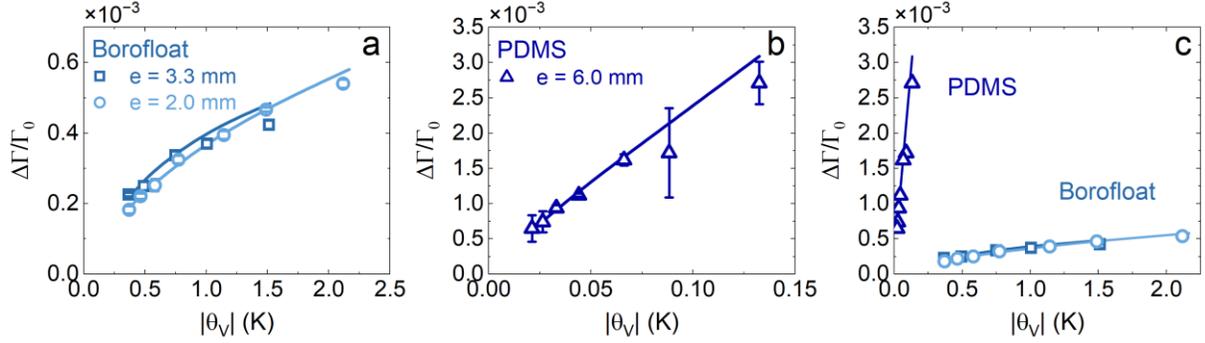

*Figure 5. (a) Variation of transmittance as a function of the modulus of the mean temperature $|\theta_V|$ for thick layers of Borofloat (e = 3.3 mm, squares and e = 2 mm, circles) (b) The same data for PDMS (e = 6,0 mm, triangles). (c) A comparison between thermotransmittance measured in thick Borofloat and PDMS.*

*Table 2. Thermo optical coefficients measured for PDMS and Borofloat glass. For thick layers, $\kappa_R$ and $\kappa_A$ are the estimated coefficients presented in Figure 5.*

| Materials | Water | PDMS | Borofloat glass | |
|---|---|---|---|---|
| $\lambda_{probe}$ (µm) | $5.33 \pm 0.25$ | $4.26 \pm 0.1$ | $3.25 \pm 0.5$ | |
| $\mu_0$ (mm)$^{-1}$ | 12.68 | 0.23 | 1.37 | |
| $e$ (mm) | 0.1 | 6.0 | 2.0 | 3.3 |
| $\kappa_R$ (K$^{-1}$) | $1.40 \times 10^{-3}$ | $1.64 \times 10^{-1}$ | $4.08 \times 10^{-4}$ | $3.64 \times 10^{-4}$ |
| $\kappa_A$ ((K.mm)$^{-1}$) | $9.57 \times 10^{-2}$ | $-4.44 \times 10^{-2}$ | $-1.70 \times 10^{-4}$ | $-1.60 \times 10^{-4}$ |

In this work, we have developed a method to measure the thermo-optic coefficients in thermally thick materials that are semitransparent in the mid-IR spectral range. This novel approach extends thermo-optic coefficient measurements beyond thermally thin materials in which the temperature is considered homogeneous throughout the thickness. Whereas in thin materials the variation of transmittance is directly proportional to the variation of temperature, such that $\frac{\Delta\Gamma}{\Gamma_0} = \kappa \Delta T$, the general case that includes thermally thick samples relies on distinguishing the interface temperatures and the temperature averaged throughout the thickness.

We used a 1D thermal model to validate our measurements, first on thin samples consisting of thin layers of water sandwiched between optically transparent glass, then on a thick layer of Borofloat, for which we obtained the same thermo-optic coefficients as those measured with a different method in our previous work, and finally on a thick layer of PDMS. The thermo-optic coefficients in PDMS were found to be more than two orders of magnitude higher than in the two other materials, which enables measurements of small temperature variations. We have also observed the transition from thermally thick to thermally thin material by changing the pump laser modulation frequency. Here, we were limited in frequency by our broadband IR source, the probe, as achieving higher frequencies implies a rapidly decreasing intensity and therefore lower SNR. Nonetheless, the method remains valid independently of the frequency and this implementation opens perspectives for thinner layers in which boundary conditions can be tuned by changing the pump frequency.

This technique is also particularly well suited to microfluidic chips, as the three materials investigated here are key components. More broadly, it is applicable for temperature mapping in multi-layer materials, 3D temperature field reconstruction, as well as buried heat source identification. In particular, we have demonstrated that our method is sensitive to temperature increases as low as 30 mK in PDMS.

**Acknowledgements**

The authors gratefully acknowledge the French National Research Agency (ANR) for its support through the project No. ANR-22-CE50-0015, the Nouvelle-Aquitaine Region (Project No. AAPR2021-2020-12035510), and the Reseau BEST Impulsion, "Industry of the Future," of the University of Bordeaux. The authors would like to thank the entire Imaging and Thermal Characterization team at the I2M laboratory for their support and the insightful discussions shared throughout this work.